\documentclass[a4paper,twoside,11pt]{article} 

\usepackage[english]{babel} %
\usepackage{lmodern}
\usepackage[T1]{fontenc}
\usepackage[latin9]{inputenc}	

\newlength{\defaultparindent}
\usepackage{latexsym}
\usepackage{amsmath}
\usepackage{amsfonts}
\usepackage{amsthm}
\usepackage{bbold}
\usepackage{multirow}
\usepackage{hyperref}
\hypersetup{colorlinks=true,citecolor=green,linkcolor= blue}
\usepackage{todonotes} 
\usepackage{soul} 
\usepackage{mathtools} 
\usepackage{rotating} 
\usepackage{arydshln} 

\usepackage[arXiv]{optional} 
\def\mynote{\todo} 

\opt{AACA}{
\def\cal{\mathcal}
}

\opt{x,AACA}{
\input{mbh_defs_TeX}
}

\opt{arXiv,std,JMP,JOPA}{

\newtheorem{MS_Proposition}{Proposition}

\def\ie{i.e.\ }
\def\eg{e.g.\ }
\def\myisom{\cong} 
\newcommand{\R}{\ensuremath{\mathbb{R}}} 
\newcommand{\C}{\ensuremath{\mathbb{C}}} 
\newcommand{\F}{\ensuremath{\mathbb{F}}} 
\newcommand{\Z}{\ensuremath{\mathbb{Z}}} 
\newcommand{\HH}{\ensuremath{\mathbb{H}}} 
\newcommand{\Identity}{\ensuremath{\mathbb{1}}} 
\def\my_span#1{\mbox{Span}\left(#1\right)} 

\def\bino#1#2{\left( \! \begin{array}{c} #1 \\ #2 \end{array} \! \right)}
\def\dotinformula{\;\; \mathrm{.}} 

\newcommand{\comm}[2]{\ensuremath{\left[ #1, #2 \right]}}
\newcommand{\anticomm}[2]{\ensuremath{\left\{ #1, #2 \right\}}} 
\def\Cl{{\cal C}\ell}	
\newcommand{\myCl}[3]{\ensuremath{{{\cal C}\ell} {\left( #1, #2 \right)}}}	
\newcommand{\myClf}[3]{\ensuremath{{{\cal C}\ell}_{#3} {\left( #1, #2 \right)}}}	

}

\opt{JMP}{
}

\def\h_eigen{\eta}
\def\g_eigen{\theta}
\def\mygen{e} 
\def\mydual{{*}} 

\newcommand{\Mod}[1]{\ (\textup{mod}\ #1)}
\def\mymod{\Mod} 
\def\mypb{\overset{.}{+}} 

\begin{document}

\opt{x,std,arXiv,JMP,JOPA}{
\title{{\bf
On Clifford Algebras and binary integers 
} 
	}

\author{\\
	\bf{Marco Budinich}%
%
%
\\
	University of Trieste and INFN, Trieste, Italy\\
	\texttt{mbh@ts.infn.it}\\
%
%
%
	}
\date{ \today }
\maketitle
}

\opt{AACA}{
\title[On Spinors of Zero Nullity]{On Spinors of Zero Nullity}

\author{Marco Budinich}
\address{Dipartimento di Fisica\\
	Università di Trieste \& INFN\\
	Via Valerio 2, I - 34127 Trieste, Italy}
\email{mbh@ts.infn.it}
}

\begin{abstract}
We show that the binary representation of the integers has a role to play in many aspects of Clifford algebras.
\end{abstract}


\opt{AACA}{
\keywords{Clifford algebra, spinors, Fock basis.}
\maketitle
}

\section{Introduction}
\label{Introduction}
\opt{margin_notes}{\mynote{mbh.note: for paper material see log pp. 640 ff. and 650 ff.}}
In 1913 {\'{E}}lie Cartan introduced spinors \cite{Cartan_1913, Cartan_1937} and, after more than a century, this
source is still pouring. Spinors were later thoroughly investigated by Claude Chevalley \cite{Chevalley_1954} in the mathematical frame of Clifford algebras where they were identified as elements of minimal left ideals of the algebra.

Clifford algebras are a remarkably powerful tool to deal with automorphisms of linear (vector) spaces. By%
\opt{margin_notes}{\mynote{mbh.note: Dieudonné extended Cartan theorem in "Sur les groupes classiques. Paris: Hermann. 1948"}}
{} Cartan-Dieudonné theorem \cite{Cartan_1937} all these automorphisms can be represented by a succession of reflections (emergence of Pin and Spin groups hierarchies). Each reflection is in turn essentially a binary process represented by one bit of information: multiply a given coordinate by $1$ or $-1$. So Clifford algebras are interwoven with reflections that share some properties with binary operations.

A second aspect of Clifford algebras in which one can feel the presence of binary numbers is the Witt decomposition of its vector space: it is well known that the the anticommutation relations fulfilled by the null vectors of the Witt basis are identical to those satisfied by the annihilation and creation operators of states subject to Fermi statistics. This property, first observed by Brauer and Weyl in 1935 can be used to construct the Fock basis of spinor spaces \cite{BudinichP_1989} and we will show that each element of this basis induces two binary signatures that, interpreted as integer numbers, represent row and column indices in the isomorphic matrix algebra.

\opt{margin_notes}{\mynote{mbh.note: in Lounesto book (history at the end) it's written that is Cartan that first discovered the periodicities that are frequently ascribed to Bott!}}
The third appearance of binary numbers comes from ``periodicity properties'' of Clifford algebras: in particular for the real Clifford algebra of vector space $V = \R^{k,l}$ the value of $k-l \pmod{8}$ determines uniquely both the underlying division algebra and to which of the 8 classes of the Brauer-Wall group of $\R$ the Clifford algebra belongs \cite[p.~89]{BudinichP_1988e}, on the other hand the symmetry properties of the invariant bilinear forms of spinor space depend on $k+l \pmod{8}$ and both values give rise to the ``spinorial chessboard'', that gives the title to the book \cite[p.~109]{BudinichP_1988e}. We note that if the numbers $k+l, k-l \pmod{8}$ are written in binary form they have just 3 bits.

%
%
%

All this said it is sensible to assume that the binary representation of integers could have a role to play in Clifford algebras. We will show that it is actually so: in this paper we explore the binary numbers that surfaces in different spots of Clifford algebras under a unified point of view.

In section~\ref{Binary_representations_integers} we start by some remarks on binary numbers; in section~\ref{Clifford_algebra_and_EFB} we show which is the exact meaning of the binary representation of the matrix indices of Clifford algebras. In last section~\ref{Classification_real_Clifford} we show that there is something to learn from the binary expression of the $(k,l) \pmod{8}$ signature of the vector space and we exploit it to present a new graphical representation of the $k - l$ periodicity property of Clifford algebras.

For the convenience of the reader we tried to make this paper as elementary and self-contained as possible.

\section{Binary representation of the integers}
\label{Binary_representations_integers}
\opt{margin_notes}{\mynote{mbh.ref: for this part see pp. 650 ff.}}
We resume some simple facts on the binary representation of an integer $n$:
\begin{equation}
\label{binary_expansion}
n = \left\{
\begin{array}{l l}
\sum_{i = 0}^\infty n_i 2^i & n_i \in \{ 0, 1 \} \\
\sum_{i = 0}^\infty \frac{1 - \prescript{}{i}n}{2} 2^i & \prescript{}{i}n \in \{ 1, -1 \}, \; \prescript{}{i}n = 1 - 2 n_i
\end{array} \right.
\end{equation}
and the bits $n_i$ and $\prescript{}{i}n$ of the binary expansion form a group, respectively with addition modulo 2 and multiplication, isomorphic to $\Z_2$; we will switch between the two forms as and when it suits to us. It is easy to see that
$$
\begin{array}{l l l}
n_i & = & \frac{n \mymod{2^{i + 1}} - n \mymod{2^{i}}}{2^i} \\
\prescript{}{i}n & = & (-1)^{\left\lfloor\frac{n}{2^i}\right\rfloor}
\end{array}
$$
the second formula descending from the observation that the binary representation of $\left\lfloor{n}/{2^i}\right\rfloor$ is equal to that of $n$, right shifted of $i$ bits.

Representing an integer $n$ with a finite number $k$ of bits one is de facto implementing modular arithmetic, namely $n \mymod{2^k}$. In this case the customary ``2-complement'' representation for negative numbers is given by $-n = 2^k - n$, that satisfies $n + (-n) \equiv 0 \mymod{2^k}$. By Clifford algebra periodicities many algebra properties depend on integers modulo 8, thus meaning that just the three least significant bits of integer $n$ are relevant so that we will frequently use $n' := n \mymod{8}$ and it is simple to verify that, if $n' \ne 0$ then $- n \mymod{8} = 8 - n'$.

We just remind the mechanics of bit addition, corresponding to logical XOR function,
$$
0 \mypb x = x \qquad 1 \mypb x = \bar{x} \qquad x \mypb x = 0 \qquad x \mypb \bar{x} = 1 \qquad x \in \{ 0, 1 \}
$$
where the bar represents the ``opposite'' bit and the operator $\mypb$ represents addition modulo 2.

A remarkable result that can be derived from Lucas' theorem (see \eg \cite{Granville_1997}) is
\begin{equation}
\label{Lucas_theorem}
\prescript{}{i}n = (-1)^{\bino{n}{2^i}}
\end{equation}
that shows that the formulas $(-1)^{\frac{n (n - 1)}{2}}$ and $(-1)^n$, that frequently occur in Clifford algebras, are just the two least significant bits $\prescript{}{1}n$ and $\prescript{}{0}n$ of the binary representation of $n$.

\section{Clifford algebra and matrix indices}
\label{Clifford_algebra_and_EFB}
We show that the row and column indices of the matrix algebra, isomorphic to a Clifford algebra, are already present in the algebra in \emph{binary form}.

We consider the classical case of neutral vector spaces: $V = \C^{2 m}$ or $\R^{m, m}$, spaces which Witt decomposition is the direct sum of two totally null (isotropic) subspaces of dimension $m$, and we exploit the properties of the Extended Fock Basis (EFB) \cite{BudinichM_2009, Budinich_2011_EFB} that we summarize in the sequel; a more exhaustive version of this summary appears in \cite{Budinich_2016}.

Let $V$ be our neutral vector space over $\F = \C$ or $\R$; any base $\mygen_1, \mygen_2, \ldots, \mygen_{n}$ with $n = 2 m$ generates the Clifford algebra $\myCl{m}{m}{}$ that results: simple, central and isomorphic to a matrix algebra: $\myCl{m}{m}{} \myisom \F(2^m)$. We can always choose $\mygen_i$'s such that
\opt{margin_notes}{\mynote{mbh.note: remember that $\myCl{2}{2}{} \myisom \myCl{3}{1}{} \myisom \R(4)$ (Dirac algebra ?), while $\myCl{1}{3}{} \myisom \HH(2)$ and for both the even subalgebra $\myisom \C(2)$ ...}}
$$
2 \, \mygen_i \cdot \mygen_j = \mygen_i \mygen_j + \mygen_j \mygen_i := \anticomm{\mygen_i}{\mygen_j} = 2 \delta_{i j} (-1)^{i+1}
$$
while $\anticomm{\mygen^i}{\mygen_j} = 2 \delta^i_j$ and 
\opt{margin_notes}{\mynote{mbh.note: This result is proved exactly in Porteous book 1981 648 Theorem~13.27, p.~ 249.}}
\opt{margin_notes}{\mynote{mbh.ref: For $\C$ the ``signature'' can be freely chosen, see \eg p. 496' and Math\_643.}}
\begin{equation}
\label{space_signature}
\left\{ \begin{array}{l l l}
\mygen_{2 i - 1}^2 & = & 1 \\
\mygen_{2 i}^2 & = & -1
\end{array} \right.
\qquad i = 1,\ldots,m \dotinformula
\end{equation}
The Witt, or null, basis of the vector space $V$ is defined, for both fields:
\begin{equation}
\label{formula_Witt_basis}
\left\{ \begin{array}{l l l}
p_{i} & = & \frac{1}{2} \left( \mygen_{2i-1} + \mygen_{2i} \right) \\
q_{i} & = & \frac{1}{2} \left( \mygen_{2i-1} - \mygen_{2i} \right)
\end{array} \right.
\Rightarrow
\left\{\begin{array}{l l l}
\mygen_{2i-1} & = & p_{i} + q_{i} \\
\mygen_{2i} & = & p_{i} - q_{i}
\end{array} \right.
\quad i = 1,2, \ldots, m
\end{equation}
that, with $\mygen_{i} \mygen_{j} = - \mygen_{j} \mygen_{i}$, gives
\begin{equation}
\label{formula_Witt_basis_properties}
\anticomm{p_{i}}{p_{j}} = \anticomm{q_{i}}{q_{j}} = 0
\qquad
\anticomm{p_{i}}{q_{j}} = \delta_{i j}
\end{equation}
showing that all $p_i, q_i$ are mutually orthogonal, also to themselves, that implies $p_i^2 = q_i^2 = 0$, at the origin of the name ``null'' given to these vectors.

Following Chevalley we define spinors as elements of a minimal left ideal; \emph{simple} (pure) spinors are those spinors that are annihilated by a null subspace of $V$ of maximal dimension $m$.

\bigskip

The EFB of \myCl{m}{m}{} is given by the $2^{2 m}$ different sequences
\begin{equation}
\label{EFB_def}
\psi_1 \psi_2 \cdots \psi_m := \Psi \qquad \psi_i \in \{ q_i p_i, p_i q_i, p_i, q_i \} \qquad i = 1,\ldots,m
\end{equation}
in which each $\psi_i$ can take four different values and we reserve $\Psi$ for EFB elements and $\psi_i$ for its components. The main characteristics of EFB is that all its $2^{2 m}$ elements $\Psi$ are simple spinors. In a nutshell the EFB extends to the entire algebra the Fock basis of its spinor spaces \cite{BudinichP_1989}.

\bigskip

We start observing that $\mygen_{2 i - 1} \mygen_{2 i} = q_i p_i - p_i q_i := \comm{q_i}{p_i}$ and that for $i \ne j$ $\comm{q_i}{p_i} \psi_j = \psi_j \comm{q_i}{p_i}$. With (\ref{formula_Witt_basis_properties}) and (\ref{EFB_def}) it is easy to calculate
\begin{equation}
\label{commutator_property}
\comm{q_i}{p_i} \psi_i = h_i \psi_i \qquad h_i = \left\{
\begin{array}{l l}
+1 & \quad \mbox{iff $\psi_i = q_i p_i \; \mbox{or} \; q_i$} \\
-1 & \quad \mbox{iff $\psi_i = p_i q_i \; \mbox{or} \; p_i$}
\end{array} \right.
\end{equation}
and the value of $h_i$ depends on the first null vector appearing in $\psi_i$. We have thus proved that $\comm{q_i}{p_i} \Psi = h_i \Psi$ and thus each EFB element $\Psi$ defines a vector $h = (h_1, h_2, \ldots, h_m) \in \{ \pm 1 \}^m$ called ``$h$ signature''. In EFB the identity $\Identity$ and the volume element
\begin{equation}
\label{omega_def}
\omega = \mygen_1 \mygen_2 \cdots \mygen_{n}
\end{equation}
(scalar and pseudoscalar) assume similar expressions:
\begin{equation}
\label{identity_omega}
\begin{array}{l l l}
\Identity & = & \anticomm{q_1}{p_1} \anticomm{q_2}{p_2} \cdots \anticomm{q_m}{p_m} \\
\omega & = & \comm{q_1}{p_1} \comm{q_2}{p_2} \cdots \comm{q_m}{p_m}
\end{array}
\end{equation}
with which
\begin{equation}
\label{Weyl_eigenvector}
\omega \Psi = \h_eigen \; \Psi \qquad \h_eigen := \prod_{i = 1}^m h_i = \pm 1 \dotinformula
\end{equation}
Each EFB element $\Psi$ has thus an eigenvalue $\h_eigen$: the {\em chirality}. Similarly the ``$g$ signature'' of an EFB element is the vector $g = (g_1, g_2, \ldots, g_m) \in \{ \pm 1 \}^m$ where $g_i$ is the parity of $\psi_i$ under the main algebra automorphism $\alpha(\mygen_i) = - \mygen_i$. With this definition and with (\ref{commutator_property}) we easily obtain
\begin{equation}
\label{commutator_left_property}
\psi_i \comm{q_i}{p_i} = g_i \comm{q_i}{p_i} \psi_i = h_i g_i \psi_i
\end{equation}
and thus
\begin{equation}
\label{EFB_are_left_Weyl}
\Psi \; \omega = \h_eigen \g_eigen \; \Psi \qquad \h_eigen \g_eigen = \pm 1 \qquad \g_eigen := \prod_{i = 1}^m g_i
\end{equation}
where the eigenvalue $\h_eigen \g_eigen$ is the product of chirality times $\g_eigen$, the global parity of the EFB element $\Psi$ under the main algebra automorphism. We can resume saying that all EFB elements are not only Weyl eigenvectors, \ie right eigenvectors of $\omega$ (\ref{Weyl_eigenvector}), but also its left eigenvectors (\ref{EFB_are_left_Weyl}) with respective eigenvalues $\h_eigen$ and $\h_eigen \g_eigen$.

$h$ and $g$ signatures play a crucial role in this description of \myCl{m}{m}{}: it is easy to see that any EFB element $\Psi = \psi_1 \psi_2 \cdots \psi_m$ is uniquely identified by its $h$ and $g$ signatures: $h_i$ determines the first null vector ($q_i$ or $p_i$) appearing in $\psi_i$ and $g_i$ determines if $\psi_i$ is even or odd, see (\ref{EFB_def}). Moreover it can be shown \cite{Budinich_2011_EFB} that \myCl{m}{m}{}, as a vector space, is the direct sum of $2^m$ subspaces of different $h$ signatures and also of $2^m$ subspaces of different $h \circ g$ signatures, where $h \circ g \in \{ \pm 1 \}^m$ is the Hadamard (entrywise) product of $h$ and $g$ signature vectors; $h \circ g = (h_1 g_1, \ldots, h_m g_m)$.

We can thus uniquely identify each of the $2^{2 m}$ EFB elements with these two ``indices'' \ie
\begin{equation}
\label{EFB_index_def}
\Psi_{a b} \left\{
\begin{array}{l l}
a \in \{ \pm 1 \}^m \quad \mbox{is the} \quad h \mbox{ signature} \\
b \in \{ \pm 1 \}^m \quad \mbox{is the} \quad h \circ g \mbox{ signature}
\end{array} \right.
\end{equation}
so that the generic element of $\mu \in \myCl{m}{m}{}$ can be written as $\mu = \sum_{a b} \xi_{a b} \Psi_{a b}$ with $\xi_{a b} \in \F$ and one can prove \cite{Budinich_2011_EFB} that:
\begin{equation}
\label{EFB_products}
\Psi_{a b} \Psi_{c d} = s(a,b,d) \, \delta_{b c} \Psi_{a d} \qquad s(a,b,d) = \pm 1
\end{equation}
where $\delta_{b c}$ is $1$ if and only if the two signatures $b$ and $c$ are equal and the sign $s(a,b,d)$, slightly tedious to calculate, depends on the indices; in \cite{Budinich_2011_EFB} it is shown how it can be calculated recursively. With this result we can easily calculate the most general Clifford product
\begin{equation*}
\mu \nu = \left(\sum_{a b} \xi_{a b} \Psi_{a b} \right) \left(\sum_{c d} \zeta_{c d} \Psi_{c d} \right) = \sum_{a d} \rho_{a d} \Psi_{a d} \qquad \rho_{a d} = \sum_{b} s(a,b,d) \xi_{a b} \zeta_{b d}
\end{equation*}
that shows that EFB elements naturally display a matrix structure, it also emerges that different $h \circ g$ signatures identify different minimal left ideals and thus different spinor spaces of the algebra.

This matrix structure is mirrored in the isomorphic matrix algebra $\F( 2^m )$ where $a$ and $b$ are nothing else than the row and column indices of $\Psi_{a b}$ when interpreted as binary numbers using the second representation of (\ref{binary_expansion}). We illustrate this structure with an example taken from \cite{Budinich_2011_EFB}: we give the EFB for $\Cl(2,2) \myisom \myCl{3}{1}{} \myisom \R(4)$ with $h$ (rows) and $h \circ g$ (columns) signatures; the number in parenthesis is the conversion of the binary form
\begin{equation}
\label{EFB_Cl_2_2_example}
\bordermatrix{& ++ (0) & +- (1) & -+ (2) & -- (3) \cr
++ (0) & q_1 p_1 \, q_2 p_2 & q_1 p_1 \, q_2 & q_1 \, q_2 p_2 & q_1 \, q_2 \cr
+- (1) & q_1 p_1 \, p_2 & q_1 p_1 \, p_2 q_2 & - q_1 \, p_2 & - q_1 \, p_2 q_2 \cr
-+ (2) & p_1 \, q_2 p_2 & p_1 \, q_2 & p_1 q_1 \, q_2 p_2 & p_1 q_1 \, q_2 \cr
-- (3) & - p_1 \, p_2 & - p_1 \, p_2 q_2 & p_1 q_1 \, p_2 & p_1 q_1 \, p_2 q_2 \cr }
\end{equation}
and the signs of matrix elements come from (\ref{EFB_products}).

We stress that EFB constitutes a base of the algebra itself and not of its representations and the matrix formalism, with row and column indices, emerges right from the algebra. As a side remark we observe that this formulation provides the faster algorithm for actual Clifford product evaluations \cite{BudinichM_2009} resulting a factor $2^m$ faster than algorithms based on gamma matrices.

We have shown that for neutral spaces matrix multiplication rules are integral part of Clifford algebra without the need to resort to representations and that the $h$ and $h \circ g$ signatures, responsible of chirality and parity of spinors, are just the binary form of the matrix indices.

\section{Periodicities of Clifford algebras}
\label{Classification_real_Clifford}
%
%
\begin{table}
\centering
\begin{tabular}{c | c | c c c c c c c c |}
\multicolumn{2}{c}{} & \multicolumn{8}{c}{$\nu$} \\
\cline{3-10}
\multicolumn{1}{c}{} & & $0$ & $1$ & $2$ & $3$ & $4$& $5$& $6$& $7$ \\
\cline{2-10}
\multirow{8}{*}{$n$}
%
%
%
%
& $0$ & $\R$ & & & \multicolumn{1}{r :}{} & & & & \\
& $1$ & & $2 \R$ & & \multicolumn{1}{r :}{} & & & & \\
& $2$ & $\R(2)$ & & $\R(2)$ & \multicolumn{1}{r :}{} & & & & \\
& $3$ & & $2 \R(2)$ & & \multicolumn{1}{r :}{$\C(2)$} & & & & \\
\cdashline{3-10}
& $4$ & $\R(4)$ & & $\R(4)$ & \multicolumn{1}{r :}{} & $\HH(2)$ & & & \\
& $5$ & & $2 \R(4)$ & & \multicolumn{1}{r :}{$\C(4)$} & & $2 \HH(2)$ & & \\
& $6$ & $\R(8)$ & & $\R(8)$ & \multicolumn{1}{r :}{} & $\HH(4)$ & & $\HH(4)$ & \\
& $7$ & & $2 \R(8)$ & & \multicolumn{1}{r :}{$\C(8)$} & & $2 \HH(4)$ & & $\C(8)$ \\
\cline{2-10}
\end{tabular}
\caption{The real, universal, Clifford algebras of $V = \R^{k,l}$ depending on $n = k + l$ and $\nu = k - l$. $\R(2)$ is the matrix algebra of real $2 \times 2$ matrices and $2 \R$ stands for $\R \oplus \R$; for negative values of $\nu$ the division algebra can be derived remembering that, for modulo 8 values, $- \nu = 8 - \nu$.}
\label{n_nu_table_reduced}
\end{table}
Let $n$ be the dimension of the vector space $V$: for a real vector space of signature $(k,l)$, $n = k + l$ and we define $\nu := k - l$ (also called the index of the inner product \cite[p.~33]{BudinichP_1988e}). Clearly $n$ and $\nu$ give a bijective transformation of the $k, l$ plane. In the literature it is customary to arrange the periodicity properties of Clifford algebras in a table of $n$ vs $\nu$, like in table~\ref{n_nu_table_reduced}, or $- \nu$ like in the the spinorial clock of \cite[p.~122]{BudinichP_1988e}%
%
%
%
%
\footnote{ in the literature there are several different ways of organizing this information; for example in the book of Porteous \cite[Table~13.26]{Porteous_1981} there is a double twist with respect to our convention: he defines the real vector space as $\R^{k,l}$ with $k$ indicating \emph{timelike} coordinates but defines Clifford algebra so that $x^2 = - x \cdot x$ for $x \in \R^{k,l}$ so that at the end the Clifford algebra is the same and in the table $n = k + l$ runs downward but his $\nu = - k + l$ is minus our one. In the same table of Wikipedia entry ``\href{https://en.wikipedia.org/wiki/Classification_of_Clifford_algebras}{\underline{Classification of Clifford algebras}}'' the situation is exactly like ours but the $\nu = k - l$ direction goes to the left so, at a first glance, the table results equal to Porteous'! Finally in the the spinorial clock \cite[p.~122]{BudinichP_1988e} the clockwise direction is given by $- \nu \mymod{8}$ because only with $- \nu$ one obtains that for a given Clifford algebra the ``preceding'' algebra is also its even subalgebra.}%
.
\opt{margin_notes}{\mynote{mbh.ref: About the Babel of ...tables see log. p. 640'}}

In what follows we show that it is also instructive to arrange the various kinds of Clifford algebras as depending on the bits of $\nu$ (more precisely the division algebra of the matrix algebra of Wedderburn-Artin theorem). To add a meaning to these bits we briefly reproduce some standard results: given any vector space $V$ and its volume element (\ref{omega_def}) it is immediate to derive
$$
\omega \mygen_i = (-1)^{(n - 1)} \mygen_i \omega \qquad 1 \le i \le n
$$
that may be put at work to prove:
\begin{figure}
\centering
\includegraphics[scale=2]{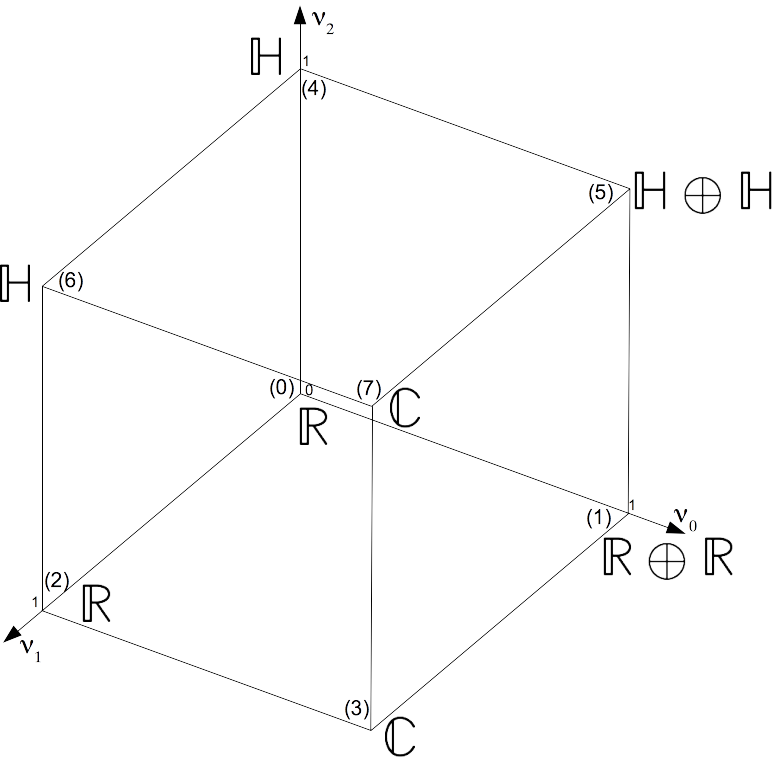}
\caption{The real Clifford algebras of $V = \R^{k,l}$ depending from the three least significant bits of the binary representation of $\nu$, the number in parenthesis near the cube vertex.}
\label{nu_drawing}
\end{figure}
\begin{MS_Proposition}
\label{n_even_iff_central}
A Clifford algebra over fields $\R$ or $\C$ is central if and only if the dimension of the vector space $n$ is even. If $n$ is odd $\omega$ belongs to the center of the algebra (that thus is not central).
\end{MS_Proposition}
We remark that in the real case $n$ even implies that the algebra is also simple but not vice versa: for $n$ odd there are simple algebras \eg $\myClf{0}{1}{\R} \myisom \C$. For complex vector spaces $n$ even is in one to one correspondence with a central \emph{and} simple algebra. Anyhow in both cases, when $n$ is even (and also $\nu$ since $\nu = n - 2 l$) we can conclude that the algebra is central and simple.

For Clifford algebras over a real vector space $V = \R^{k, l}$ it is a standard exercise to calculate $\omega^2$ and using (\ref{Lucas_theorem})
$$
\omega^2 = (-1)^{\frac{(k - l)(k - l - 1)}{2}} = (-1)^{\bino{\nu}{2}} = \prescript{}{1}\nu
$$
namely the second bit of the integer $\nu = k - l$. This holds also for Euclidean and complex spaces when $k - l = k = n$ and $\omega^2 = \prescript{}{1}n$, the second bit of $n$.

\opt{margin_notes}{\mynote{mbh.note: division ring or algebra ? Artin-Wedderburn theorem uses rings; but by Frobenius theorem $\R, \C$ and $\HH$ are division algebras so I put algebra}}
Finally from table~\ref{n_nu_table_reduced} we realize that the structure of division algebras for $0 \le \nu \le 3$ is repeated exactly, with a factor $2^4$ in dimensions over $\R$, to $4 \le \nu \le 7$ if we replace the occurrences of $\R$ with $\HH$; they are respectively upper left and lower right dashed rectangles in table~\ref{n_nu_table_reduced}. This implies that the third bit $\nu_2$ indicates whether the underlying division algebra is $\R$ or $\HH$.

It is evocative to replace table~\ref{n_nu_table_reduced} with a 3-dimensional cube with the 3 axes corresponding respectively to $\nu_0, \nu_1$ and $\nu_2$ and with each division algebra on its corresponding vertex as done in figure~\ref{nu_drawing}. We see that the cube face $\nu_0 = 0$ ($\nu, n$ even) contains all central simple algebras whereas the face $\nu_0 = 1$ ($\nu, n$ odd) contains all non central algebras. The two faces $\nu_1 = 0, 1$ correspond respectively to $\omega^2 = 1, -1$ and finally the faces $\nu_2 = 0, 1$ contain respectively $\R$ or $\HH$ division algebras.

The vertical edge $\nu_0 = \nu_1 = 1$ contains the cases in which $\omega$ belongs to the center and $\omega^2 = -1$ namely in these cases the center is $\{ \Identity, \omega \}$ that is isomorphic to $\C$. The other edge $\nu_0 = 1, \nu_1 = 0$ contains the cases in which the center is again $\{ \Identity, \omega \}$ but $\omega^2 = 1$ namely the cases of ``double'' algebra $\R \oplus \R$ and $\HH \oplus \HH$.

The advantage of this representation is that each axis of the cube represent a well defined property of the algebra, thus superseding traditional $\nu \pmod{4}, \nu \pmod{2}$ characterizations, and the properties of Clifford algebras result easier to visualize.

\bigskip

Given the binary value of $\nu$ we can learn something about the Clifford algebra but this is not the entire story. Indeed there are other periodicities of the Clifford algebra, that give rise to the spinorial chessboard \cite[p.~109]{BudinichP_1988e} where the chessboard has cartesian coordinates $k, l \pmod{8}$. Clearly given the value of $\nu$ different possible alternatives for $n$ remain.

These further ``degrees of freedom'' are responsible for the spinorial chessboard and appears in a different form producing the eight double coverings of the group O, the Dabrowski groups \cite{Dabrowski_1988}. Dabrowski defines three further bits, named $a, b, c$, that completely determines the characteristics of Clifford algebra; his work has been subsequently developed by Varlamov \cite{Varlamov_2001}.

We show that this uncertainty can be removed, and the bits $a, b, c$ determined, from the values of the three least significant bits of $n$ that explain both the Dabrowski groups and the spinorial chessboard.

\bigskip

To do this we briefly review the fundamental automorphisms of Clifford algebras that are responsible for the values of these bits. In general in Clifford algebras there are four automorphisms corresponding to the two involutions and to the two antinvolutions induced by the orthogonal involutions $\Identity_V$ and $-\Identity_V$ of vector space $V$ \cite[Theorem~13.31]{Porteous_1981}. They are called fundamental or discrete automorphisms and under composition form a finite group, isomorphic to $\Z_2 \otimes \Z_2$ \cite{Varlamov_2001}.

In what follows we restrain to the simpler case of $n$ even leaving the odd case for future analysis. For $n$ even we have seen that the Clifford algebra is central simple and, by Skolem -- Noether theorem, all its automorphisms are inner. In \cite{Budinich_2016} it is shown that the inner elements $\omega$ (\ref{omega_def}) and
\begin{equation}
\label{tau_def}
\tau := \left\{
\begin{array}{l l}
\mygen_{k+1} \mygen_{k + 2} \cdots \mygen_{k + l} & \quad \mbox{for} \; k,l \; \mbox{even} \\
\mygen_{1} \mygen_{2} \cdots \mygen_{k} & \quad \mbox{for} \; k,l \; \mbox{odd.}
\end{array} \right. 
\end{equation}
(here $\mygen_i^2 = 1$ for $i \le k$ and $\mygen_i^2 = -1$ for $i > k$) give, together with $\omega \tau$:
\begin{eqnarray*}
\omega \mygen_i \omega^{-1} & = & - \mygen_i \qquad \qquad \forall \: 1 \le i \le n \\
\tau \mygen_{i} \tau^{-1} & = & \mygen_{i}^{\mydual} = \mygen_{i}^{-1} \\
\omega \tau \mygen_{i} (\omega \tau)^{-1} & = & - \mygen_{i}^{\mydual}
\end{eqnarray*}
where $\mygen_{i}^{\mydual}$ is the dual of $\mygen_{i}$. Together with identity, they define the 4 fundamental involutions of Clifford algebra (not to be mistaken with antinvolutions, \eg reversion)%
\footnote{ Substantially $\tau$ is the inner, real, version of $B$ of \cite{BudinichP_1988e, Trautman_1998}; in a nutshell $B: S \to S^{\mydual}$ intertwines the equivalent \emph{complex representations} $\gamma, \gamma^{\mydual}$ of a real Clifford algebra in dual complex spaces $S$ and $S^{\mydual}$ and gives also the automorphism $\gamma(\mygen_{i})^{\mydual} = B \gamma(\mygen_{i}) B^{-1}$ and, for even $n = 2m$, $B^{\mydual} = (-1)^{\frac{m (m - 1)}{2}} B$. On the other hand $\tau$ is the inner element of the real algebra giving automorphism $\mygen_{i}^{\mydual} = \tau \mygen_{i} \tau^{-1}$; moreover given its definition (\ref{tau_def}) it is easy to see that $\tau^{\mydual} = \tau^{-1} = \tau^2 \tau$ (whereas $B^{-1} : S^{\mydual} \to S$ is different from $B^{\mydual}: S \to S^{\mydual}$). To compare $B^{\mydual}$ with $\tau^{\mydual}$ the correct thing is to compare $\tau^2$ (\ref{tau_squared}) with $(-1)^{\frac{m (m - 1)}{2}}$.

The complex representations of Clifford algebra of $\R^{k,l}$ are obtained from the restriction of the complex representations of Clifford algebra of $\C^{k+l}$ and, in the even dimensional case $k+l=n=2m$, this last Clifford algebra is isomorphic to that of matrices $\C(2^m)$. Restricted to the real case these matrices give a complex representation of the real Clifford algebra of $\R^{m,m}$ that, with (\ref{tau_squared}), shows that there is no mismatch between $B^{\mydual}$ and $\tau^{\mydual}$ since they coincide for the complex representations of the real algebra.}%
. We show that the bits $\tau^2 = \pm 1$ and $(\omega \tau)^2 = \pm 1$, together with those of $\nu$, are enough to fully determine the Clifford algebra eliminating all uncertainties.

\opt{margin_notes}{\mynote{mbh.ref: log p. 646}}
From $\tau$ definition (\ref{tau_def}) it is simple to obtain
\begin{equation}
\label{tau_squared}
\tau^2 = \left\{
\begin{array}{l l}
(-1)^{\frac{l (l - 1)}{2}} & \quad \mbox{for} \; k,l \; \mbox{even} \\
(-1)^{\frac{k (k - 1)}{2}} & \quad \mbox{for} \; k,l \; \mbox{odd}
\end{array} \right. \\
\end{equation}
and since $\omega \tau = \tau \omega$ for $k,l$ even while $\omega \tau = - \tau \omega$ for $k,l$ odd we have
\begin{equation}
\label{omega_tau_squared}
(\omega \tau)^2 = \left\{
\begin{array}{l l}
\omega^2 \tau^2 & \quad \mbox{for} \; k,l \; \mbox{even} \\
- \omega^2 \tau^2 & \quad \mbox{for} \; k,l \; \mbox{odd.}
\end{array} \right. \\
\end{equation}

\begin{MS_Proposition}
\label{n_bits_from_tau}
For $n$ even its three least significant bits, $\prescript{}{2}n \prescript{}{1}n \prescript{}{0}n$, are:
\begin{eqnarray*}
\prescript{}{2}n & = & \prescript{}{2}\nu \tau^2 \\
\prescript{}{1}n & = & (\omega \tau)^2 \tau^2 \\
\prescript{}{0}n & = & 1
\end{eqnarray*}
\end{MS_Proposition}
\begin{proof}
Let the binary representations of $k,l \mymod{8}$ be respectively $k_2 k_1 k_0$ and $l_2 l_1 l_0$, since $n = k + l$ is even, necessarily $n_0 = \nu_0 = 0$ and thus $k_0 = l_0$. We prove the relation for $\prescript{}{2}n$ calculating $\prescript{}{2}n \prescript{}{2}\nu$ in the form $n_2 \mypb \nu_2$; we examine separately two cases: let first $k, l$ be even and thus $k_0 = l_0 = 0$; in this case we write the relation $v + 2 l = n$ in binary form and limited to the 3 least significant bits (\ie $\pmod{8}$, remember also that in binary form $2 l$ is just $l$ shifted to the left by one bit)
\begin{equation}
\label{n_nu_1}
\nu_2 \nu_1 0 + l_1 0 0 = n_2 n_1 0
\end{equation}
\opt{margin_notes}{\mynote{mbh.note: to prove these relations one needs to check that the ``carries'' are $c_0 = c_1 = 0$.}}
from which, since there are no carries in the sum of the two rightmost bits, we derive $n_2 \mypb \nu_2 = l_1$ and since for $k,l$ even, by (\ref{tau_squared}) and (\ref{Lucas_theorem}), $\tau^2 = \prescript{}{1}l$ we can conclude that in this case the thesis is true. The second case is $k, l$ odd and thus $k_0 = l_0 = 1$; in this case we use the relation $\nu + n = 2k$
\begin{equation}
\label{n_nu_2}
\nu_2 \nu_1 0 + n_2 n_1 0 = k_1 1 0
\end{equation}
and we derive (again no carries) $n_2 \mypb \nu_2 = k_1$ and since for $k,l$ odd by (\ref{tau_squared}) $\tau^2 = \prescript{}{1}k$ also in this case the thesis is true.

To prove the result for the second bit $\prescript{}{1}n$ we get from (\ref{omega_tau_squared})
\begin{equation*}
(\omega \tau)^2 \tau^2 = \left\{
\begin{array}{l l l l}
\omega^2 & = & \prescript{}{1}\nu & \quad \mbox{for} \; k,l \; \mbox{even} \\
- \omega^2 & = & - \prescript{}{1}\nu & \quad \mbox{for} \; k,l \; \mbox{odd.}
\end{array} \right. \\
\end{equation*}
and by (\ref{n_nu_1}) we see that in the first case $\prescript{}{1}n = \prescript{}{1}\nu$ while in the second case by (\ref{n_nu_2}) we derive $\prescript{}{1}n = - \prescript{}{1}\nu$; the proposition is thus proved also for $\prescript{}{1}n$.
\end{proof}

We have thus proved that knowing $\nu$ and the value of the two bits $\tau^2$ and $(\omega \tau)^2$ we can determine also $n \mymod{8}$ and thus remove all periodicities from Clifford algebra. In Varlamov's notation \cite{Varlamov_2001} $(a, b, c) = ((\omega \tau)^2, \tau^2, \omega^2) = (\prescript{}{1}n, \prescript{}{2}\nu \prescript{}{2}n, \prescript{}{1}\nu)$.

An interesting remark is that knowing the characteristics of the algebra and the values of $\omega^2, \tau^2, (\omega \tau)^2$ we can get the space signature $(k, l) \pmod{8}$.

\section{Conclusions}
\label{Conclusions}
We have presented the role played by binary numbers in different aspects of Clifford algebra. In certain respects Clifford algebras may appear as a bridge joining continuous and discrete worlds, these findings add another connection between them. We remark that the complete mosaic has several missing tiles that need to be put in place, among them:
\begin{itemize}
\item the matrix indices coming from EFB have to be extended to other real signatures and to complex odd dimensional spaces,
\item $\tau$ form and properties have to be extended to odd dimensional spaces,
\item the determination of $n \mymod{8}$ from the bits $\tau^2$ and $(\omega \tau)^2$ has to be extended to the case of odd dimensional spaces.
\end{itemize}


\opt{x,std,AACA}{
\bibliographystyle{plain} 

\bibliography{mbh}
}

\opt{arXiv,JMP}{
%
%

%
%
}

\opt{final_notes}{
\newpage

\section*{Things to do, notes, etc.......}

\noindent General things to do:
\begin{itemize}
\item check again the $B / \tau$ puzzle, log. p. 651;
\item ...
\end{itemize}

\subsection*{Old parts no more needed}
When adding two integers $k + l = n$ in binary format the general rule is
$$
n_i \equiv k_i + l_i + c_{i - 1} \mymod{2} := k_i \mypb l_i \mypb c_{i - 1}
$$
where $c_{i - 1}$ is the possible ``carry'' generated in the sum of the previous bit and we define the operator $\mypb$ for addition modulo 2; only for the least significant bit $n_0 = k_0 \mypb l_0$ and $n_0$ depends only on the corresponding bits of the two addends.

\bigskip

Moreover by previous observation
\begin{equation}
\label{Lucas_theorem_gen}
\prescript{}{(i + k)}n = (-1)^{\bino{\left\lfloor\frac{n}{2^k}\right\rfloor}{2^i}} \dotinformula
\end{equation}

\bigskip

We tackle first the simpler case of a Clifford algebra over the complex field $\C$: here there exists only one possible ``variable'', $n$ and it is sufficient to know $n \mymod{4} = n_1 n_0$: $n_0$ signals if the algebra is central simple or not while we already saw that $\omega^2 = \prescript{}{1}n$; the four possibilities fully determine the corresponding Clifford algebra and its fundamental automorphism group: in particular $n_0$ determines if the algebra is simple or semisimple and $\omega^2$ defines, for each case, if in the fundamental automorphism group all the elements commute ($\omega^2 = 1$) or anticommute ($\omega^2 = -1$) \cite{Varlamov_2001}.
\opt{margin_notes}{\mynote{mbh.ref: \cite{Varlamov_2001} pages 8 and 31--32}}

\bigskip

%
%
\begin{sidewaystable} 
\centering
\begin{tabular}{c | c | c c c c c c c c c c c c c c c |}
\multicolumn{2}{c}{} & \multicolumn{15}{c}{$\nu$} \\
\cline{3-17}
\multicolumn{1}{c}{} & & $-7$ &$-6$ &$-5$ &$-4$ & $-3$ & $-2$ & $-1$ & $0$ & $1$ & $2$ & $3$ & $4$& $5$& $6$& $7$ \\
\cline{2-17}
\multirow{8}{*}{$n$}
%
%
%
%
& $0$ & & & & & & & & $\R$ & & & & & & & \\
& $1$ & & & & & & & $\C$ & & $2 \R$ & & & & & & \\
& $2$ & & & & & & $\HH$ & & $\R(2)$ & & $\R(2)$ & & & & & \\
& $3$ & & & & & $2 \HH$ & & $\C(2)$ & & $2 \R(2)$ & & $\C(2)$ & & & & \\
& $4$ & & & & $\HH(2)$ & & $\HH(2)$ & & $\R(4)$ & & $\R(4)$ & & $\HH(2)$ & & & \\
& $5$ & & & $\C(4)$ & & $2 \HH(2)$ & & $\C(4)$ & & $2 \R(4)$ & & $\C(4)$ & & $2 \HH(2)$ & & \\
& $6$ & & $\R(8)$ & & $\HH(4)$ & & $\HH(4)$ & & $\R(8)$ & & $\R(8)$ & & $\HH(4)$ & & $\HH(4)$ & \\
& $7$ & $2 \R(8)$ & & $\C(8)$ & & $2 \HH(4)$ & & $\C(8)$ & & $2 \R(8)$ & & $\C(8)$ & & $2 \HH(4)$ & & $\C(8)$ \\
\cline{2-17}
\end{tabular}
\caption{The real, universal, Clifford algebras of $V = \R^{k,l}$ depending on $n = k + l$ and $\nu = k - l$. $\R(2)$ is the matrix algebra of real $2 \times 2$ matrices and $2 \R$ stands for $\R \oplus \R$.} 
\label{n_nu_table}
\end{sidewaystable} 

} 

\end{document}